\documentclass[conference]{IEEEtran}
\usepackage{graphicx}
\usepackage{refstyle}
\usepackage{cite}

\begin{document}

\title{Fuzzy Membership Function Implementation \\with Memristor}
\author{
\IEEEauthorblockN{Azamat Marlen}
\IEEEauthorblockA{School of Electrical and\\Computer Engineering\\
Nazarbayev University\\
Astana, Kazakhstan\\
Email: azamat.marlen@nu.edu.kz}

\and
\IEEEauthorblockN{Anuar Dorzhigulov}
\IEEEauthorblockA{School of Electrical and\\Computer Engineering\\
Nazarbayev University\\
Astana, Kazakhstan\\
Email: adorzhigulov@nu.edu.kz}
}

\maketitle

\begin{abstract}

The neuro-fuzzy system is network which resemble human-like operation of the naturally imprecise data and decision-making. This paper proposes implementation of the fundamental module of the neuro-fuzzy system - membership function (MF), realized as a analog electronic hardware. The memristive crossbar arrays are used as the architecture for proposed MF analog circuit.  The main advantages of the memristive crossbar circuit are size, energy efficiency and fault tolerance compared to another analog alternatives. The conducted crossbar SPICE simulation with MS model of the memristor results confirm the performance and highlighted benefits of the proposed circuit.

\end{abstract}


\section{Introduction}

Current technological development produces huge demand of intelligent systems capable of autonomous big data processing. Neuro-fuzzy systems, which are based on fuzzy logic and neural networks, have wide range of applications today \cite{art1}. In fact, the mentioned systems have several advantages and the most significant among them is natural capability to work with imprecise and linguistic data, mimicking human way of the data processing and decision making.

For a long time, most of fuzzy inference systems were implemented as digital devices and solutions \cite{in1}. However, this approach limits the development of the systems that requires highly parallel operation \cite{in2}. In order to fully explore the advantages of the neuro-fuzzy systems, it requires dedicated high speed and parallel hardware.

Multiple researches propose conventional CMOS transistors based circuits for signal fuzzification \cite{art4,6210400,VALIZADEHYAGHMOURALI2018128,6613335}. Those CMOS designs demonstrate considerable performance: high output shape control, speed and robustness, efficiency in power and area. However, such membership function generation (MFG) circuits requires external memory to memorize the MF parameters, such as slope, centering and width. As a result, there is a need for additional circuitry for memory and interfacing. Utilization of conventional digital memory architectures and ACD/DAC for interfacing could become a major bottleneck in the performance of the neuro-fuzzy system in a whole, especially if large scale system is desired.

Memristor, the novel fundamental circuit element predicted by Leon Chua in 1971 \cite{art2}, is a viable solution for a design problem of the fully analog hardware with desired characteristics \cite{Memristors_book}. Application of memristor is predominantly based on its plasticity. Memristor is capable to change and passively remember its current state until the next change occurs, just like synapses in human brain. Moreover, memristor can be conveniently be operated for writing, erasing and reading the states. Such property can be utilized for a passive analog memory. MF is used for input fuzzification. In other words it should give a certain shape to the input range, which is typically triangular, trapezoidal or bell. The shape parameters then can be easily stored as a memristor states. 

Currently, such application of the memristors is gaining an interest among hardware designers for machine learning and neural network architectures \cite{Irmanova2018,8023844,Smagulova2018}. It has been shown, that memristor as a device and memristive arrays, can be used for a memory storage and as a part of the processing architecture. Memristor can be used to combine the memory and processing in one wholesome architecture. Moreover, some works suggest this concept for neuro-fuzzy applications \cite{6657808,art3}. The purpose of this work is to demonstrate the implementation of the MF shape generation on a fragment of the meristive crossbar array.


This paper will consist of several parts. First, general introduction to the memristor and memristive crossbar architecture. Next, result demonstration based on the SPICE circuit simulations. Finally, the discussion and concluding results.





\section{Memristor}

\begin{figure}[h] 
\centering 
\includegraphics[scale=0.95]{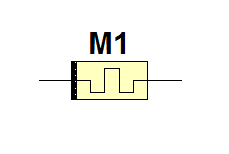} 
\caption{Simple Memristor} 
\label{fig:one} 
\end{figure}

Memristor is one of the fundamental circuit elements as resistor, capacitor and inductor. It is passive device that provides a functional relation between flux and charge, voltage and current \cite{art2}. The symbol of memristor is presented in the Fig. \ref{fig:one}.

\begin{figure}[h] 
\centering 
\includegraphics[scale=0.4]{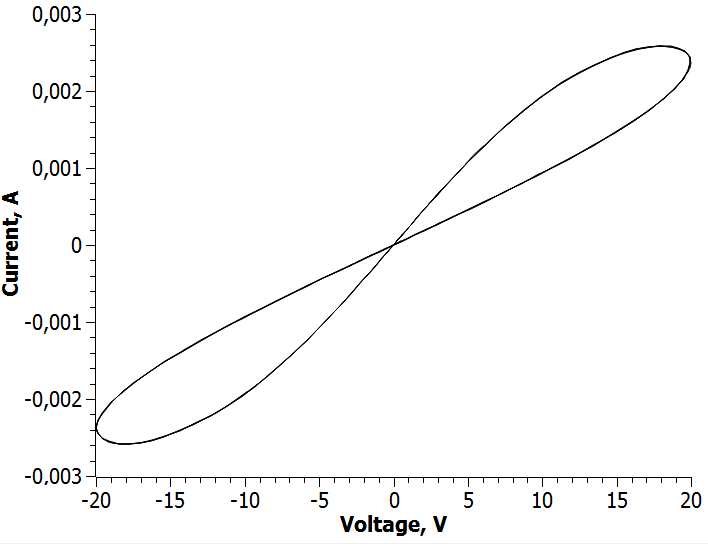} 
\caption{Hysteresis Loop} 
\label{fig:two} 
\end{figure}

The I-V characteristics of the device is represented as hysteresis loop, Fig. \ref{fig:two}. From the graph the fundamental property of the memristor can be observed - the ability to alter the I-V relation (resistance), depending on the voltage sweep amplitude.

\begin{figure}[h] 
\centering 
\includegraphics[scale=0.41]{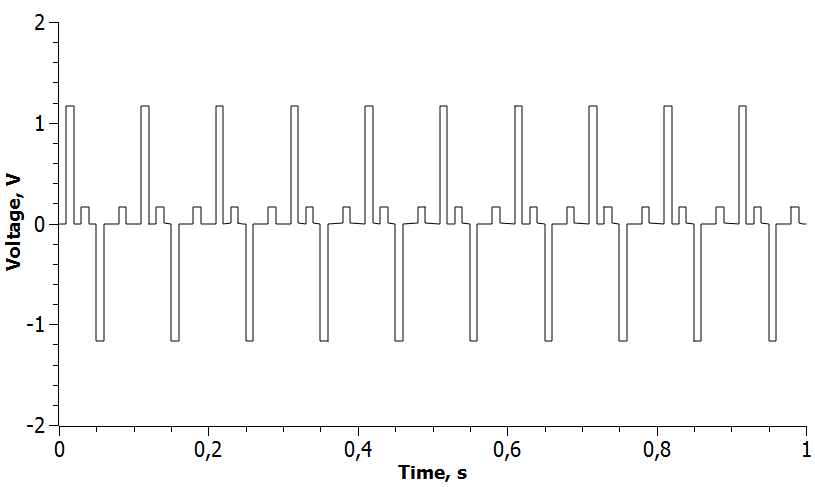} 
\caption{Voltage signal used to test the behavior of the simulated memristor model. Pulse of about 1 $V$ amplitude is used to "write" while -1 $V$ is used for "reset" the state of the memristor. Minor pulse of less than 200 $mV$ used for "reading".} 
\label{fig:three} 
\end{figure}

Two main states of the memristor can be characterized - states of low ($R_{on}$ or "write") and high resistance ($R_{off}$ "reset"). Typically, each state can be obtained by suppling certain amount of the positive or negative voltage for low and high resistance states respectively. The voltage at which resistance switch occurs called threshold voltage. A much smaller voltage in range between two threshold voltages is used to measure the resistance without risk of affecting it, Fig \ref{fig:three}.
 
The voltage and current across the memristor can be found as \cite{art3}:

\begin{equation}
V=\frac{d\phi}{dt}
\end {equation} 
\begin{equation}
i=\frac{dq}{dt}
\end{equation}

From these two equations of voltage and current and its linear dependency, it is possible to derive new quantity:

\begin{equation}
V=M(q)*i
\end{equation}
\begin{equation}
M(q)=\frac{d\phi}{dq}
\end{equation}

This new quantity is called memory resistance, or memristance, ($M(q)$) and measured in $\Omega$.

The mathematical model for the total memristance:

\begin{equation} 
M(w)=\frac{w}{D}*R_{on} + (1-\frac{w}{D})*R_{off} 
\end{equation} 
\begin{equation}
w(t)=w_0+\frac{\mu R_{on}}{D}*q(t)
\end{equation}

where $D$ is full length of memristor, $R_{on}$ is the
resistance when full length is doped and $R_{off}$ is the  resistance when total width of memristor becomes undoped , $w_0$ is the initial value of $w$, $\mu$ is the average ion mobility and $q(t)$ is the amount of electric charge \cite{art3, in3}. The value of memristance depends on voltage and current applied to it.

\begin{figure}[h] 
\centering 
\includegraphics[scale=0.5]{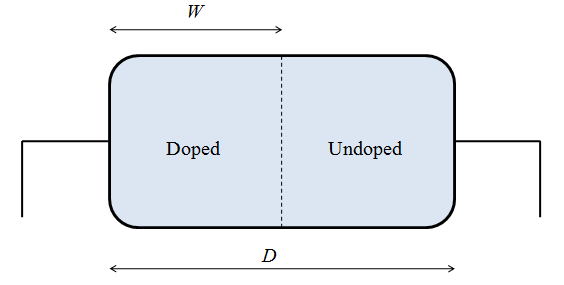} 
\caption{Structure of Memristor} 
\label{fig:four} 
\end{figure}

\section{Membership Function generation circuit}

The proposed design is based on memristor crossbar to obtain the membership function. The crossbar circuit consists of parallel wires of $N$ inputs which are crossing perpendicularly parallel wires of $M$ outputs. The points of intersections are called a crosspoints, where each input wire connected to each output wire with memristor. In 2D crossbar array, there is $N \times M$ connection nodes or neurons \cite{art2}. The interest in crossbar architecture is related to the simplicity of circuit and its fault tolerance which caused by large number connections between inputs and outputs \cite{art5}. 

\begin{figure}[h] 
\centering 
\includegraphics[scale=0.4]{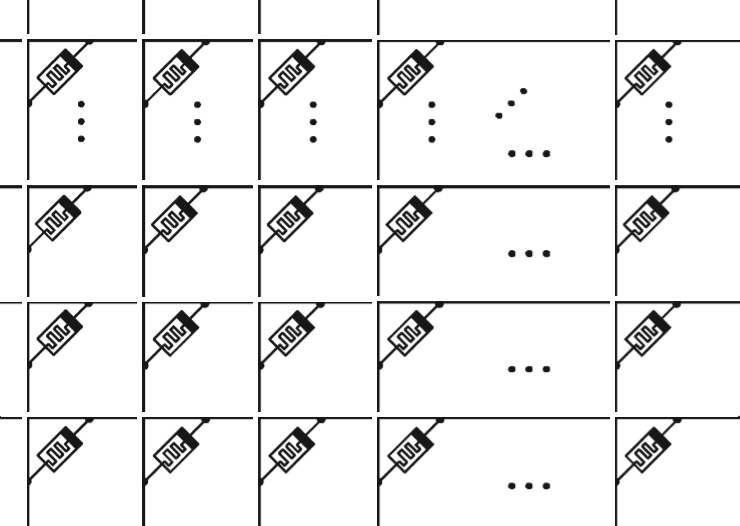} 
\caption{Memristor Crossbar-based Circuit} 
\label{fig:five} 
\end{figure} 

\begin{figure}[h] 
\centering 
\includegraphics[scale=0.7]{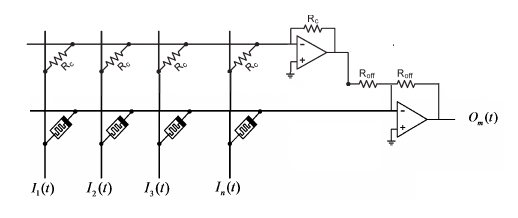} 
\caption{Memristive Crossbar Analog Circuit} 
\label{fig:six} 
\end{figure}

This design allows to construct discrete fuzzification of the input, by dividing each input into multiple voltage amplitude levels, that are connected to the crossbar array separately, Fig. \ref{fig:six}. 

The several number of resistor is connected to circuit. The resistor  of special resistance $R_{off}$ which shows the highest of memristance at undoped region  is connected to the operational amplifier as feedback. As the result of connection memristances with resistances, the summing circuit based on operational amplifiers was obtained at the analog circuit. One more type of resistor is operating in the circuit. The second row of crossbar is shown as parallel connection of resistors $R_{c}$ that creates the same summing circuit as previous. It can be counted as additional input of crossbar at amplifiers. The aim of the resistors $R_{c}$ is to detect the threshold voltage that affects to the memristance \cite{art8}.


The crossbar design implies usage of various memristance in each neuron. This result can be achieved by different ways. Some researchers state that the voltage higher than threshold is producing gradually increasing or decreasing of resistance \cite{art6}, \cite{art7}. The other types of memristor models operating in different way. Some of them is capable to decrease the resistance at each pulse of negative threshold voltage, but reset to $R_{off}$ at positive value of threshold voltage, whereas some of memristors is not gradual at all.


\begin{figure}[h] 
\centering 
\includegraphics[scale=0.55]{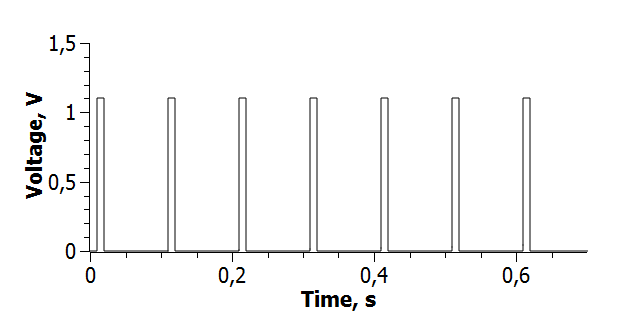} 
\caption{Example of the voltage pulses that used to alter the memristance of the device.} 
\label{fig:seven} 
\end{figure}

Figure \ref{fig:seven} shows the example of applied switching voltage to one of the neurons to achieve unique memristance. The width of pulse and its number of cycles affect to the memristance value \cite{art6}, \cite{art7}.


\section{Simulation}

For the simulation, the MS memristor model in crossbar of 8 neurons has been simulated in SPICE. The threshold voltages ($V_{th}$) is approximately 1 and -1 $V$. The lower level resistance ($R_{on}$) is approximately 3 k$\Omega$ and resistance at higher level ($R_{off}$) is about 62 k$\Omega$. 

\begin{figure}[h] 
\centering 
    \includegraphics[scale=0.5]{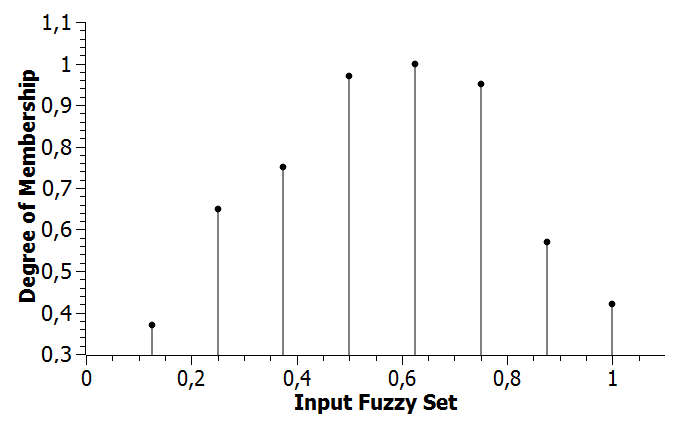} 
\caption{Discrete membership function obtained from memristive $8\times1$ crossbar array simulation} 
\label{fig:eight}
\end{figure}

For the first neuron, 3 cycles of pulse signal with period of 100 milliseconds voltage was applied. The values of output and input voltages were obtained as the result of simulation. The same operation with constant voltage value and period were done to another 7 neurons. The number of cycles were chosen randomly and were 7, 8, 12, 15, 11, 6 and 4 for each subsequent input respectively, and results of the simulation can be seen on a Fig. \ref{fig:eight}. As it can be seen, the discrete input fuzzification is achievable.
 
\section{Discussion}
%

Proposed circuit is capable to produce discrete shape of the MF. The memristance can be finely tuned with control voltages to obtain any desired shape, be it conventional triangular, trapezoidal or Gaussian bell. The resolution of the output shape is limited only by the number of the memristors used to represent the fuzzification of the desired input. The results show the simulation for just the one bar within the whole 2D array. It means, all other bars can be used to construct multiple independent MF shapes for a single input. Such feature may be useful for a neuro-fuzzy systems where multiple MFs are obtained from a single input, for example ANFIS. Moreover, crossbar array structure can be considered to be implemented as a 3D architecture \cite{7763751}, providing a great prospects for large scale, yet compact realization of the possible neural systems in a hardware.


More possible solution might be found by more deep research in the topic of fuzzy systems and membership function. The project topic needs further investigation to improve the way of implementation of fuzzy membership function by memristor.

\section{Conclusion}
This paper was illustrated how the membership function of any shape can be implemented by memristor crossbar circuit. The method of implementation of fuzzy membership function that had been discussed is based on memristors in a crossbar array architectures. The advantage of that analog circuit is simplicity and fault tolerance. The crossbar-based circuit's computational methodology is similar to human brain  and easily can be implemented in the field of artificial intelligence. 

\bibliographystyle{IEEEtran} 
\bibliography{references}

\begin{thebibliography}{10}
\providecommand{\url}[1]{#1}
\csname url@samestyle\endcsname
\providecommand{\newblock}{\relax}
\providecommand{\bibinfo}[2]{#2}
\providecommand{\BIBentrySTDinterwordspacing}{\spaceskip=0pt\relax}
\providecommand{\BIBentryALTinterwordstretchfactor}{4}
\providecommand{\BIBentryALTinterwordspacing}{\spaceskip=\fontdimen2\font plus
\BIBentryALTinterwordstretchfactor\fontdimen3\font minus
  \fontdimen4\font\relax}
\providecommand{\BIBforeignlanguage}[2]{{%
\expandafter\ifx\csname l@#1\endcsname\relax
\typeout{** WARNING: IEEEtran.bst: No hyphenation pattern has been}%
\typeout{** loaded for the language `#1'. Using the pattern for}%
\typeout{** the default language instead.}%
\else
\language=\csname l@#1\endcsname
\fi
#2}}
\providecommand{\BIBdecl}{\relax}
\BIBdecl

\bibitem{art1}
M.~A. Boyacioglu and D.~Avci, ``An adaptive network-based fuzzy inference
  system (anfis) for the prediction of stock market return: the case of the
  istanbul stock exchange,'' \emph{Expert Systems with Applications}, vol.~37,
  no.~12, pp. 7908--7912, 2010.

\bibitem{in1}
S.~S{\'a}nchez-Solano, A.~Barriga, C.~Jim{\'e}nez, and J.~Huertas, ``Design and
  application of digital fuzzy controllers,'' in \emph{Fuzzy Systems, 1997.,
  Proceedings of the Sixth IEEE International Conference on}, vol.~2.\hskip 1em
  plus 0.5em minus 0.4em\relax IEEE, 1997, pp. 869--874.

\bibitem{in2}
F.~Merrikh-Bayat, S.~B. Shouraki, and F.~Merrikh-Bayat, ``Memristor
  crossbar-based hardware implementation of fuzzy membership functions,'' in
  \emph{Fuzzy Systems and Knowledge Discovery (FSKD), 2011 Eighth International
  Conference on}, vol.~1.\hskip 1em plus 0.5em minus 0.4em\relax IEEE, 2011,
  pp. 645--649.

\bibitem{art4}
T.~Kettner, C.~Heite, and K.~Schumacher, ``Analog cmos realization of fuzzy
  logic membership functions,'' \emph{IEEE Journal of Solid-state circuits},
  vol.~28, no.~7, pp. 857--861, 1993.

\bibitem{6210400}
J.~Oh, S.~Lee, and H.~J. Yoo, ``1.2-mw online learning mixed-mode intelligent
  inference engine for low-power real-time object recognition processor,''
  \emph{IEEE Transactions on Very Large Scale Integration (VLSI) Systems},
  vol.~21, no.~5, pp. 921--933, May 2013.

\bibitem{VALIZADEHYAGHMOURALI2018128}
\BIBentryALTinterwordspacing
Y.~V. Yaghmourali, A.~Fathi, M.~Hassanzadazar, A.~Khoei, and K.~Hadidi, ``A
  low-power, fully programmable membership function generator using both
  transconductance and current modes,'' \emph{Fuzzy Sets and Systems}, vol.
  337, pp. 128 -- 142, 2018, theme: Applications. [Online]. Available:
  \url{http://www.sciencedirect.com/science/article/pii/S0165011417301070}
\BIBentrySTDinterwordspacing

\bibitem{6613335}
A.~Abolhasani, M.~Tohidi, M.~Mousazadeh, A.~Khoei, and K.~Hadidi, ``A high
  speed and fully tunable mfg with new programmable cmos ota and new min
  circuit,'' in \emph{Proceedings of the 20th International Conference Mixed
  Design of Integrated Circuits and Systems - MIXDES 2013}, June 2013, pp.
  169--173.

\bibitem{art2}
D.~B. Strukov, G.~S. Snider, D.~R. Stewart, and R.~S. Williams, ``The missing
  memristor found,'' \emph{nature}, vol. 453, no. 7191, p.~80, 2008.

\bibitem{Memristors_book}
A.~James, \emph{Memristor and Memristive Neural Networks}.\hskip 1em plus 0.5em
  minus 0.4em\relax Intech, 2018.

\bibitem{Irmanova2018}
\BIBentryALTinterwordspacing
A.~Irmanova and A.~P. James, ``Neuron inspired data encoding memristive
  multi-level memory cell,'' \emph{Analog Integrated Circuits and Signal
  Processing}, Mar 2018. [Online]. Available:
  \url{https://doi.org/10.1007/s10470-018-1155-z}
\BIBentrySTDinterwordspacing

\bibitem{8023844}
O.~Krestinskaya, T.~Ibrayev, and A.~P. James, ``Hierarchical temporal memory
  features with memristor logic circuits for pattern recognition,'' \emph{IEEE
  Transactions on Computer-Aided Design of Integrated Circuits and Systems},
  pp. 1--1, 2017.

\bibitem{Smagulova2018}
\BIBentryALTinterwordspacing
K.~Smagulova, O.~Krestinskaya, and A.~P. James, ``A memristor-based long short
  term memory circuit,'' \emph{Analog Integrated Circuits and Signal
  Processing}, Apr 2018. [Online]. Available:
  \url{https://doi.org/10.1007/s10470-018-1180-y}
\BIBentrySTDinterwordspacing

\bibitem{6657808}
F.~Merrikh-Bayat, F.~Merrikh-Bayat, and S.~B. Shouraki, ``The neuro-fuzzy
  computing system with the capacity of implementation on a memristor crossbar
  and optimization-free hardware training,'' \emph{IEEE Transactions on Fuzzy
  Systems}, vol.~22, no.~5, pp. 1272--1287, Oct 2014.

\bibitem{art3}
F.~Merrikh-Bayat and S.~B. Shouraki, ``Memristive neuro-fuzzy system,''
  \emph{IEEE transactions on cybernetics}, vol.~43, no.~1, pp. 269--285, 2013.

\bibitem{in3}
R.~F. Abdel-Kader and S.~M. Abuelenin, ``Memristor model based on fuzzy window
  function,'' in \emph{Fuzzy Systems (FUZZ-IEEE), 2015 IEEE International
  Conference on}.\hskip 1em plus 0.5em minus 0.4em\relax IEEE, 2015, pp. 1--5.

\bibitem{art5}
I.~Vourkas and G.~C. Sirakoulis, ``A novel design and modeling paradigm for
  memristor-based crossbar circuits,'' \emph{IEEE Transactions on
  Nanotechnology}, vol.~11, no.~6, pp. 1151--1159, 2012.

\bibitem{art8}
M.~Zidan, H.~Omran, R.~Naous, A.~Sultan, H.~Fahmy, W.~Lu, and K.~N. Salama,
  ``Single-readout high-density memristor crossbar,'' \emph{Scientific
  reports}, vol.~6, 2016.

\bibitem{art6}
K.~D. Cantley, A.~Subramaniam, H.~J. Stiegler, R.~A. Chapman, and E.~M. Vogel,
  ``Hebbian learning in spiking neural networks with nanocrystalline silicon
  tfts and memristive synapses,'' \emph{IEEE Transactions on Nanotechnology},
  vol.~10, no.~5, pp. 1066--1073, 2011.

\bibitem{art7}
S.~H. Jo, T.~Chang, I.~Ebong, B.~B. Bhadviya, P.~Mazumder, and W.~Lu,
  ``Nanoscale memristor device as synapse in neuromorphic systems,'' \emph{Nano
  letters}, vol.~10, no.~4, pp. 1297--1301, 2010.

\bibitem{7763751}
G.~C. Adam, B.~D. Hoskins, M.~Prezioso, F.~Merrikh-Bayat, B.~Chakrabarti, and
  D.~B. Strukov, ``3-d memristor crossbars for analog and neuromorphic
  computing applications,'' \emph{IEEE Transactions on Electron Devices},
  vol.~64, no.~1, pp. 312--318, Jan 2017.

\end{thebibliography}
\end{document}